\documentclass[]{spie}
\usepackage[]{graphicx}
\usepackage{amsmath}
\usepackage{url}

\title{A new method of CCD dark current correction \\
via extracting the dark information from scientific images}

\author{Bin Ma\supit{1}, Zhaohui Shang\supit{2,1}, Yi Hu\supit{1}, Qiang Liu\supit{1}, Lifan Wang\supit{3,4}, Peng Wei\supit{1}
\skiplinehalf
\supit{1}National Astronomical Observatories, Chinese Academy of Sciences, Beijing, China; \\
\supit{2}Tianjin Normal University, Tianjin, China; \\
\supit{3}Purple Mountain Observatory, Chinese Academy of Sciences, Nanjing, China; \\
\supit{4}Texas A\&M University, College Station, Texas, USA;
}

\authorinfo{Send correspondence to Bin Ma: mabin22@gmail.com}

\begin{document}
  \maketitle

\begin{abstract}

We have developed a new method to correct dark current at relatively
high temperatures for Charge-Coupled Device (CCD) images when dark
frames cannot be obtained on the telescope. For images taken with the
Antarctic Survey Telescopes (AST3) in 2012, due to the low cooling
efficiency, the median CCD temperature was -46$^\circ$C, resulting in
a high dark current level of about 3$e^-$/pix/sec, even comparable to
the sky brightness (10$e^-$/pix/sec).  If not corrected, the
nonuniformity of the dark current could even overweight the photon
noise of the sky background. However, dark frames could not be
obtained during the observing season because the camera was operated
in frame-transfer mode without a shutter, and the telescope was
unattended in winter. Here we present an alternative, but simple and
effective method to derive the dark current frame from the scientific
images. Then we can scale this dark frame to the temperature at which
the scientific images were taken, and apply the dark frame corrections
to the scientific images.  We have applied this method to the AST3
data, and demonstrated that it can reduce the noise to a level roughly
as low as the photon noise of the sky brightness, solving the high
noise problem and improving the photometric precision. This method
will also be helpful for other projects that suffer from similar
issues.

\end{abstract}


\section{INTRODUCTION}
\label{sec:intro}

Dark current in the Charge-Coupled Device (CCD) is generated by
thermal electrons within the silicon structure of the CCD, and these
cannot be distinguished from the actual image photoelectrons. The dark
current generation rate essentially depends on the temperature, and
could be reduced dramatically as the temperature drops. Thus the CCDs
are commonly cooled to $\sim$ -100$^\circ$C with liquid nitrogen,
Peltier junctions (thermoelectric coolers, TECs), or mechanical pumps
(cryo-coolers). Sometimes the total dark current level is less than 1
$e^-$ in an image and could be neglected, otherwise e.g., in the very
long exposures, the dark current still contributes measurable noises,
hence should be corrected. The standard procedures of dark correction
are to take the dark frames with the same exposure time at the
operating temperature with the shutter closed, combine them to create
a high signal-to-noise ratio (SNR) master dark frame, and then
subtract it from the scientific images.

The trio Antarctic Survey Telescopes (AST3)\cite{Cui2008, Yuan2010}
are designed for multi-band wide-field survey at Dome A, the highest
place of the Antarctic Plateau, which is considered as a candidate of
the best optical/IR and sub-mm astronomical site on earth. AST3
has a modified Schmidt system design (with a tube about half the
length as that of a traditional Schmidt) with an entrance pupil
diameter of 500  mm and the focal ratio of $f/3.73$. AST3 has 
full pointing and tracking components. AST3 is equipped with the
STA1600FT CCD camera, designed and manufactured by Semiconductor
Technology Associates, Inc., with 10560 x 10560 pixels of 9 $\mu$m,
corresponding to 1 arcsec on the focal plane of AST3. The CCD chip is
cooled by TEC to take advantage of the extremely low air temperature
at Dome A in the winter, which is about -60$^\circ$C on
average.\footnote{\url{http://aag.bao.ac.cn/weather/plot_en.php?range=year}}

The first AST3 (AST3-1) has been deployed to Dome A, Antarctica by the
28th Chinese Antarctic Research Expedition (CHINARE) team in January
2012\cite{Li2012}, and is the largest optical telescope in Antarctica
so far. AST3-1 started operation in middle March, when the polar day
ended. After the test of focusing, pointing and tracking, it began normal
observations, but unfortunately stopped to work on May 8 due to some
malfunction in the power supply system. Until then, AST3-1 had repeatedly
surveyed nearly 500 fields ($\sim 2,000\, {\rm deg^2}$) with more than
$3,000$ 60-sec exposures, and monitored the center of Large Magellanic
Cloud with $> 4,700$ frames and several fields on the Galactic disk
with $\sim 7,000$ frames. The data were retrieved by the 29th
CHINARE team in April 2013, and consisted of more than $22,000$
frames, including both test and scientific images.

However, these data suffer critical dark current issue because the low cooling efficiency resulted in a median CCD temperature of -46$^\circ$C, and a large fraction of images were even taken at around -30$^\circ$C. According to the lab tests\cite{ma2012}, the temperature of -46$^\circ$C corresponds to a dark current level of about 3 $e^-$/pix/sec, which is comparable to the sky brightness (10 $e^-$/pix/sec). Moreover, the noise caused by nonuniformity of dark current is proportional to the dark current level, while the photon noise is proportional to the square root of the photon flux. As a result, the former will overweight the latter at certain temperature and exposure time, and limit both photometric depth and precision. Therefore the dark current must be corrected well for precise photometry. However,  the standard procedure was not possible for AST3, because its camera was operated in frame-transfer mode without a shutter, and it was unattended in winter. So dark frames could not be obtained during the observing season.  In addition, dark frames taken in laboratory cannot be used either, because it is found that the patterns in the dark frames changed. In this paper, we will present a new method to derive the dark current nonuniformity from the observational images directly, and demonstrate its correction efficiency on AST3 images.

\section{METHOD}
\label{sec:method}

The signals in a scientific image contain photons from celestial
objects such as stars and galaxies, and photons from the sky
background, as well as the dark current from the CCD. The stars are
located randomly in different celestial fields, thus can be removed
via a median algorithm when combining a large number of images in
various fields. The sky brightness is quite flat spatially after
twilight, therefore can be taken as a constant. However the dark
current is nonuniform across the entire CCD chip, thus it contains a
constant term (median level) plus a space-varying term. Finally
the brightness $I$ of a frame in pixel $(x, y)$ can be written as:

	\begin{equation}
	\label{eq:brightness}
I(x, y) = S + D(T) + \Delta d(T, x, y),
	\end{equation}
where $S$ is the sky background, $D(T)$ is the median dark current
level at temperature $T$, and $\Delta d(T, x, y)$ is the deviation
from the median dark current in the pixel $(x, y)$.
 
\subsection{Calculating Dark Current}
\label{sec:calculate}

To obtain the dark current, we need both $\Delta d(T, x, y)$ and
$D(T)$, respectively. In Eq.~\ref{eq:brightness} the sum of the first two
constant terms $S$ and $D(T)$
is practically the median value of the full image $I_0$.
Supposing there are two images taken at the same exposure time and
temperature $T_0$, but with different sky brightness, they have
different constant terms but the same fluctuation term:

	\begin{equation}
	\label{eq:dualbri}
	\begin{split}
I_1(x, y) &= I_{0,1} + \Delta d(T, x, y), \\
I_2(x, y) &= I_{0,2} + \Delta d(T, x, y).
	\end{split}
	\end{equation}

\noindent
If $I_{0,2}$ is brighter, $I_1(x, y)$ can be scaled to the equivalent
median level of $I_{0,2}$ by multiplying the ratio $k \equiv
I_{0,2}/I_{0,1}$:

	\begin{equation}
	\label{eq:scale1}
I_1'(x, y) = k I_1(x, y) = I_{0,2} + k \Delta d(T, x, y).
	\end{equation}

\noindent
Then subtracting $I_2(x, y)$ from $I_1'(x, y)$ removes the constant
term, and leaves only the fluctuation term. Consequently the dark
current nonuniformity at temperature $T$ can be calculated by the
image pair:

	\begin{equation}
	\label{eq:deltadark}
\Delta d(T, x, y) = \frac{I_1'(x, y) - I_2(x, y)}{k - 1} = \frac{k I_1(x, y) - I_2(x, y)}{k - 1}.
	\end{equation}

\noindent
To be more precise, there is another nonuniform factor in the flat
correction, the nonuniform response to the light (quantum efficiency)
of CCD pixels. Then $S$ should be written as $S f(x, y) \equiv S(1 +
\Delta f(x, y)) = (I_0 - D(t))(1 + \Delta f(x, y))$, where $f(x, y)$
is the normalized flat frame, and $\Delta f$ is the deviation of $f$
from unity. Taking into account the flat's influence, the image values
in Eq.~\ref{eq:dualbri} become:

	\begin{equation}
	\label{eq:dualbri2}
	\begin{split}
I_1(x, y) &= I_{0,1} + (I_{0,1} - D(t)) \Delta f(x, y) + \Delta d(T, x, y), \\
I_2(x, y) &= I_{0,2} + (I_{0,2} - D(t)) \Delta f(x, y) + \Delta d(T, x, y).
	\end{split}
	\end{equation}
Again, $I_1(x,y)$ can be scaled to be:
	\begin{equation}
	\label{eq:scale2}
I_1'(x, y) = k I_1(x, y) = I_{0,2} + (I_{0,2} - kD(t)) \Delta f(x,y) + k \Delta d(T, x, y).
	\end{equation}
And the difference between $I_1'(x,y)$ and $I_2(x,y)$ results in:
	\begin{equation}
	\label{eq:diff2}
I_1'(x, y) - I_2(x,y) = -(k - 1)D(T)\Delta f(x,y) + (k-1)\Delta d(T, x, y).
	\end{equation}
Therefore $\Delta d(T, x, y)$ is derived instead by:
	\begin{equation}
	\label{eq:deltadark2}
\Delta d(T, x, y) = \frac{k I_1(x, y) - I_2(x, y)}{k - 1} + D(T) \Delta f(x, y).
	\end{equation}

To derive the median dark current level $D(T)$ without simultaneous
dark frames, one simple approach is to adopt the previous lab test
results, assuming that it would not vary too much during one or two
years. Another estimation via observation is to take one frame at
temperature $T_0$, then warm the CCD by several degrees, and take
another frame at this temperature with the same exposure time.
While the sky brightness keeps constant, the dark current level $D(T)$
increases by a factor of $p$ as the temperature increases. Therefore
the difference between these two frames is $(p-1) D(T)$, and the
factor $p$ can be obtained according to Sec.~\ref{sec:scaling}, so the
median dark current level $D(T)$ can be calculated. This principle could also
be applied to estimate the dark current, nevertheless it would
interrupt the observation.

Adding the space-varying term $\Delta d(T,x,y)$ to the median level
$D(T)$, we can derive the dark frame
at temperature $T_0$ and exposure time $t_0$, then we can scale it to
the temperature and exposure time that match those of the scientific images 
for applying the dark correction.

\subsection{Scaling}
\label{sec:scaling}

Dark current increases exponentially as the temperature of the CCD
increases, and for AST3 CCD it doubles for every 7.3$^\circ$C increase
in temperature \cite{ma2012}. However this relationship was obtained
in the temperature range between -80$^\circ$C and -40$^\circ$C, and we
found that this overestimated the dark current level at higher temperature
when it was applied to the correction. We developed an alternative way
to estimate the dark current ratio directly between the dark frame and
a scientific image.

There are a large number of warm pixels in the CCD, whose dark
currents are several times higher than the normal pixels, but not as
extremely high as those of hot pixels. These warm pixels' dark currents
show similar behavior when increasing with temperature and exposure
time as normal pixels. Then the difference value between a warm pixel
and its adjacent normal pixel will also increase by the same factor.
In the scientific images, because the adjacent pixels are exposed by
the identical sky brightness, the difference between the pixel values
is just their dark current difference. As a result, the ratio
between the pixel value difference of the dark frame and that of the scientific
image can be utilized to scale the dark frame at temperature $T_0$ and
exposure time $t_0$ to those of the scientific image. Also, using as
many of warm pixels could reduce the noise.

\subsection{Error}
\label{sec:error}

When estimating dark current nonuniformity by Eq.~\ref{eq:deltadark2},
the random noise is mainly contributed by the first term, while the
systematic error is dominated by the second term, which are the uncertainties of $D(x,y)$
 and $\Delta f(x,y)$, and are roughly insignificant compared with random noise. The random noise is:

	\begin{equation}
	\label{eq:noise}
\sigma_{\Delta d}^2(x, y) = (\frac{k}{k-1})^2 \sigma_{I_1}^2(x, y) + (\frac{1}{k-1})^2 \sigma_{I_2}^2(x, y) .
	\end{equation}
where $\sigma_{I_1}^2(x, y)$ and $\sigma_{I_2}^2(x, y)$ are the Poisson
noises in each pixel if readout noise is negligible, and can be
approximate to $I_{0,1}$ and $I_{0,2}$ (in unit of electron),
respectively, except for the hot pixels. Besides, taking into account
$I_{0,2} = k I_{0,1},  (k > 1)$, the noise is:

	\begin{equation}
	\label{eq:noise2}
\sigma_{\Delta d}^2(x, y) = \frac{k(k+1)}{(k-1)^2} I_{0,1} \equiv K I_{0,1},
	\end{equation}
which indicates that it decreases as $k$ increases. When $k$ is close
to $1$, the noise tends to be infinite. This makes sense because
nothing can be derived if the image pair are identical. When $k$ is
large enough, the noise is close to the Poisson noise of the image
with the lower sky brightness. In practice, due to the saturation
problem of the CCD,
$k$ cannot be too large. In addition, a bright sky background,
either during twilight or caused by the moon, suffers significant gradient
which does not satisfy the assumption of flat background. Therefore $k$
should be chosen properly. The factor $K$ in Eq.~\ref{eq:noise2}
decreases as $k$ increases, e.g. its values are 6, 3, 2.2, 1.9,
respectively, corresponding to $k$ = 2, 3, 4, 5. Therefore $k > 3$ is
less noisy enough, and it corresponds to $> 1.2$ mag brighter, and this brightness ratio is
available in observations during dark nights.
   
\section{RESULTS}
\label{sec:results}

\begin{figure}
   \begin{center}
   \begin{tabular}{c}
   \includegraphics[height=4cm]{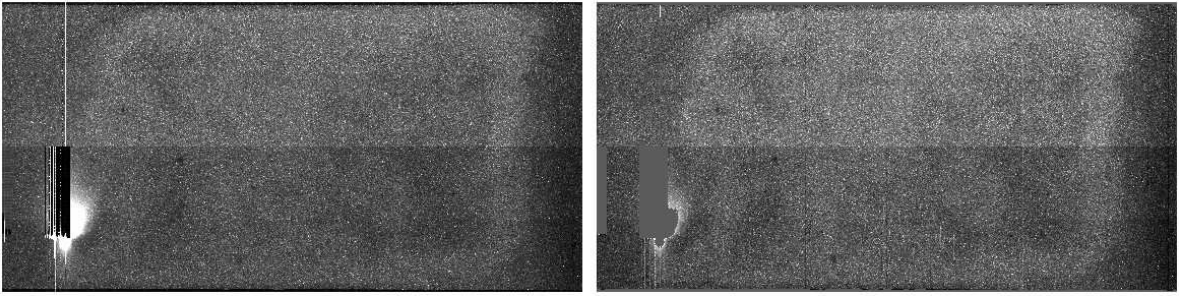}
   \end{tabular}
   \end{center}
   \caption[dark]
   {\label{fig:dark} Comparison of the lab dark frame (left) and the
derived dark frame (right).  They show similar large scale patterns.  
The black region in the bottom-left is bad pixels due 
to a damage in this engineering grade CCD.}
   \end{figure}

We have applied the method on AST3 images to correct the dark current.
We selected $230$ pairs of images, which were taken under the
temperature of $-41.9^\circ$C and with exposure time of 60 sec. We have
derived the dark current nonuniformity via Eq.~\ref{eq:deltadark2} for
each pair, and combined them with median algorithm as well as
sigma-clipping. The lower brightness images have a median background
of $\sim$ 660 ADU with the gain of 1.7 $e^-$/ADU. The brightness
ratios $k$ in these pairs are required to be greater than 3, and
finally have a median of 4.0. Therefore the random noise in one
estimation is $\sim$ 29 ADU according to Eq.~\ref{eq:noise2}. 
After median combination, the noise can be 
reduced by a factor of $\frac{1}{\sqrt{230}}$ to 1.9
ADU, which is much lower than the readout noise (7 ADU), thus would not
effect the photometric precision. The flat frame is derived by combining hundreds of 
twilight flat frames. Finally we derive the master dark
frame by adding the median dark current level from the lab test
result, and normalize it to 1 sec. In Fig.~\ref{fig:dark} we compare
this derived dark frame with lab dark frame, both of which show the
similar large scale patterns. 

\begin{figure}
   \begin{center}
   \begin{tabular}{c}
   \includegraphics[height=7cm]{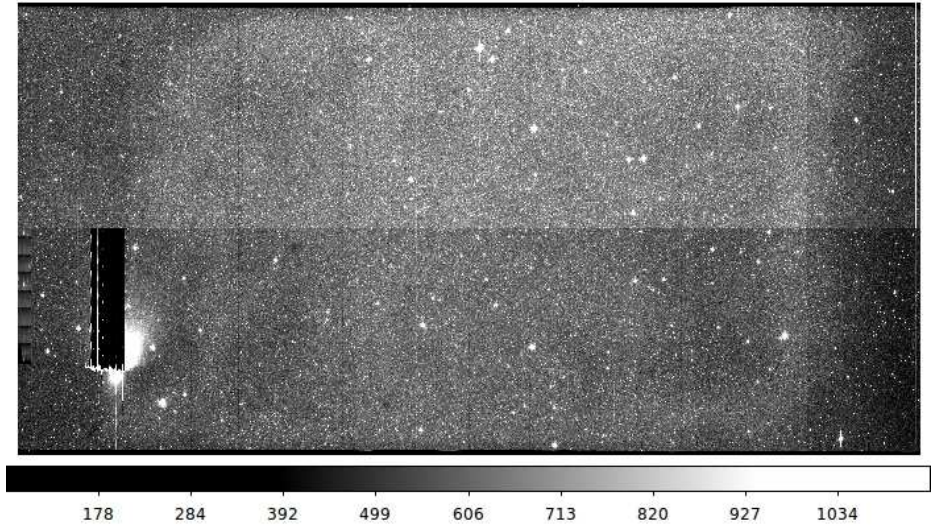}
   \end{tabular}
   \end{center}
   \caption[example]
   {\label{fig:example} Raw image $a0331.116.fit$ as an example.
The black region in the bottom-left is bad pixels due 
to a damage in this engineering grade CCD.}
   \end{figure}

\begin{figure}
   \begin{center}
   \begin{tabular}{c}
   \includegraphics[height=7cm]{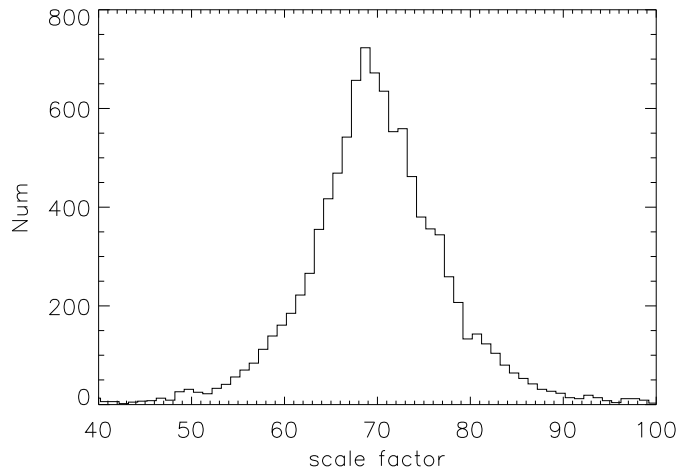}
   \end{tabular}
   \end{center}
   \caption[scale]
   {\label{fig:scale} The distribution of all scaling factors derived
using warm pixels for $a0331.116.fit$
according to Sec.~\ref{sec:scaling}.  The median value of 69.54 is
used to scale the derived master dark frame for dark correction
of the example image $a0331.116.fit$.}
   \end{figure}

\begin{figure}
   \begin{center}
   \begin{tabular}{c}
   \includegraphics[height=7cm]{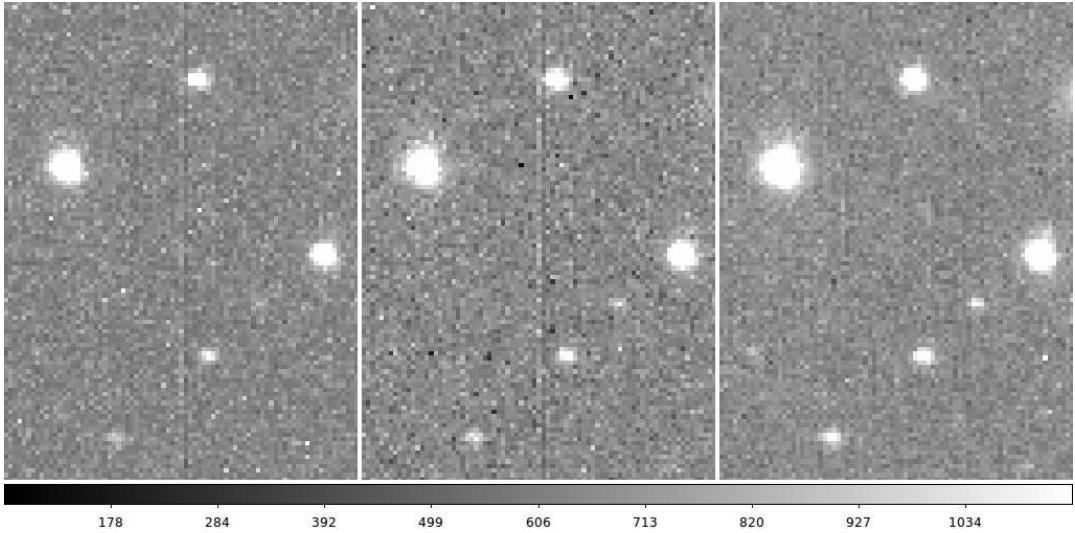}
   \end{tabular}
   \end{center}
   \caption[darkcorr]
   {\label{fig:darkcorr} A small region of the full frame is shown to
compare the dark correction effects. Left: raw image; middle:
corrected by the lab dark;
right: corrected by the derived dark. Compared with the lab dark, the
derived dark removes the hot and warm pixels much better, and improves
the SNR of the faint stars.}
   \end{figure}

\begin{figure}
   \begin{center}
   \begin{tabular}{c}
   \includegraphics[height=7cm]{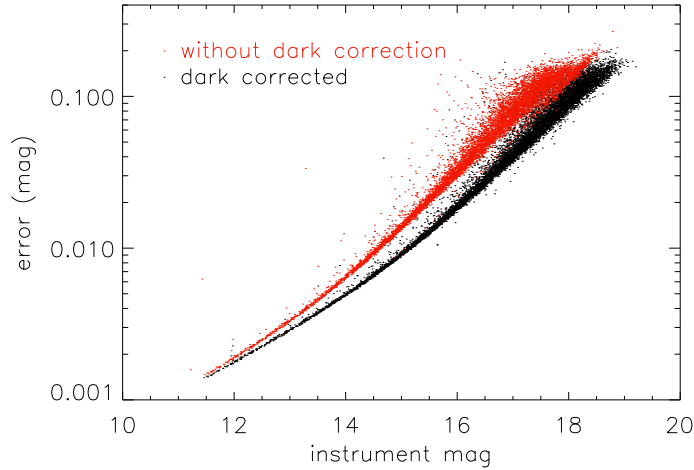}
   \end{tabular}
   \end{center}
   \caption[photdark]
   {\label{fig:photdark} The photometric error as a function of
instrument magnitude. Red dots: without dark correction; black dots:
corrected by derived dark.}
   \end{figure}

We take image $a0331.116.fit$, which was taken at temperature of
-41$^\circ$C and with exposure time of 60 sec, as an example. The raw
image is shown in Fig.~\ref{fig:example}, in which the large scale patterns
from dark current are obvious. 
We derive scaling factors using a large number of warm pixels
according to Sec.~\ref{sec:scaling} and adopt a median value of 69.54
(see Fig.~\ref{fig:scale})  to scale the derived master dark frame for dark
correction of the example image.
The correction results are illustrated in
Fig.~\ref{fig:darkcorr}. A small region of the full frame is zoomed in
to demonstrate the correction effects. In the raw image, many hot and
warm pixels are apparent. They are not corrected well by the lab dark,
that some are overcorrected while some are undercorrected,
implying that the detailed dark features have changed with time. In
contrast, these hot and warm pixels are removed much more cleanly by the derived dark,
improving the SNR of the faint stars. The background level of the raw
image in the center is 620 ADU, and its RMS of 55.4 ADU is roughly three
times of the Poisson noise (19 ADU). The background RMS is only reduced a
little to 43.5 ADU if corrected by the lab dark, while it is reduced greatly to
24.9 ADU when corrected by the derived dark. As a result, the derived dark
could improve the photometric depth by 0.9 mag, as well as the
photometric precision (see Fig.~\ref{fig:photdark}).

\begin{figure}
   \begin{center}
   \begin{tabular}{c}
   \includegraphics[height=7cm]{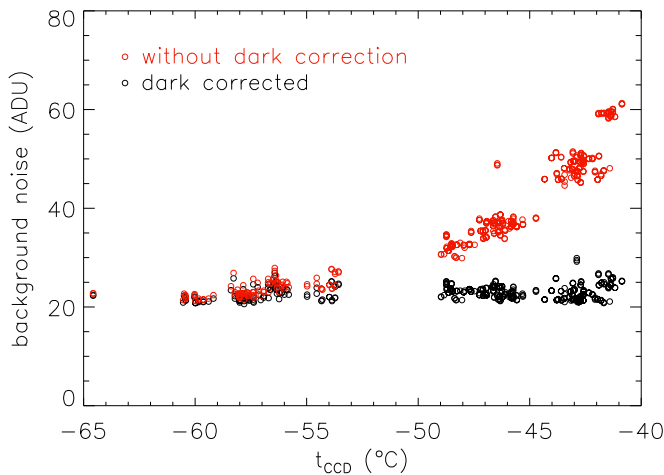}
   \end{tabular}
   \end{center}
   \caption[noise_temp]
   {\label{fig:noise_temp} The correction effect as a function of CCD
temperature. Red open circles: without dark correction; black open circles:
corrected by derived dark. The background level of these images are 500 to 600 ADU, which
means that they have very close Poisson noise.  It is clear that the
dark corrected images have much lower noise at the level of Poisson
noise.}
   \end{figure}

Furthermore, we demonstrate the correction effect as a function of CCD
temperature in Fig.~\ref{fig:noise_temp}. The background level of these
images are 500 to 600 ADU, which means that they have very close
Poisson noise level (about 18 ADU). If dark current is not corrected,
the background noise will increase significantly as the CCD temperature
rises from -55$^\circ$C. However, if the derived dark is applied for
correction, the background noises are almost constant at the Poisson
noise level even when the CCD temperature increases to -40$^\circ$C. These
results demonstrate a remarkable correction efficiency of the derived dark.

\section{CONCLUSIONS}
\label{sec:conclusion}

A new method has been developed to derive the dark current
nonuniformity from the observational images directly. The image values
are assumed to consist of the flat sky background, the median dark
current level, and the dark current nonuniformity which is
space-varying. A
image pair is chosen to have different sky background, but the same
temperature and exposure time, so that they have the unequal constant
terms, but the equal space-varying term. Scaling one image to the other
according to their median values and subtract one from the other 
could remove the constant term, and
derive the nonuniformity.  Adding to this the median dark current level from
the lab test, we can obtain a derived dark frame. Its noise approaches
the Poisson noise of the lower brightness image, and by combining a
large number of such derived dark frames from different image pairs,
we can obtain a derived master dark frame with much lower noise.
The derived master dark frame can be scaled to the same temperature and
exposure time with the scientific images for dark correction.

We have applied this method to correct the images from AST3-1 in 2012,
which suffered from relatively high CCD temperatures and no available dark frames.
We derived the master derived dark frame from 230 image
pairs. An example illustrates that after corrected by the derived dark
frame, the background noise could be reduced from 55.4 ADU to 24.9 ADU,
which is close to the Poisson noise (19 ADU). As a result, the limiting
magnitude is improved by about 0.9 mag.  Further investigation
demonstrates that the background noises, after dark correction, are almost constant at the
Poisson noise level even when the CCD temperature increases to
-40$^\circ$C.

We conclude that this method is remarkably efficient in dark
correction, and will be helpful for other projects that suffer from
similar issues like AST3. Because it is based on the observational
images only, this method does not require any extra exposures, saving
observing time
and operations. In addition, it allows one to obtain the real-time dark
current from scientific images. Although in this case, more noises are introduced by
the sky background compared with direct pure dark frames, the sky 
background noise could be reduced dramatically by combining 
huge amounts of observations and becomes lower than the readout noise, thus this
would not degrade the photometric precision.

\acknowledgments

This work has been supported by the National Basic Research Program of
China (973 Program) under grand No. 2013CB834900, the Chinese Polar
Environment Comprehensive Investigation \& Assessment Programmes under
grand No. CHINARE2014-02-03, and the National Natural Science
Foundation of China under grant No. 11003027, 11203039, and 11273019.


\bibliography{ref} 

\begin{thebibliography}{1}

\bibitem{Cui2008}
Cui, X., Yuan, X., and Gong, X., ``{Antarctic Schmidt Telescopes (AST3) for
  Dome A},'' {\em Proc. SPIE}~{\bf 7012},  70122D (2008).

\bibitem{Yuan2010}
Yuan, X., Cui, X., Gong, X., Wang, D., Yao, Z., Li, X., Wen, H., Zhang, Y.,
  Zhang, R., Xu, L., Zhou, F., Wang, L., Shang, Z., and Feng, L., ``{Progress
  of Antarctic Schmidt Telescopes (AST3) for Dome A},'' {\em Proc. SPIE}~{\bf
  7733},  77331V (2010).

\bibitem{Li2012}
{Li}, Z., {Yuan}, X., {Cui}, X., {Wang}, D., {Gong}, X., {Du}, F., {Zhang}, Y.,
  {Hu}, Y., {Wen}, H., {Li}, X., {Xu}, L., {Shang}, Z., and {Wang}, L.,
  ``{Status of the first Antarctic survey telescopes for Dome A},'' {\em Proc.
  SPIE} {\bf 8444},  84441O (2012).

\bibitem{ma2012}
Ma, B., Shang, Z., Wang, L., Boggs, K., Hu, Y., Liu, Q., Song, Q., and Xue, S.,
  ``{The test of the 10k x 10k CCD for Antarctic Survey Telescopes (AST3)},''
  {\em Proc. SPIE} {\bf 8446},  84466R (2012).

\end{thebibliography}
\bibliographystyle{spiebib}
\end{document}